\documentclass[final]{aipproc}

\layoutstyle{8x11double}

\begin{document}

\title{GRB970228 and the class of GRBs with an initial spikelike emission: do they follow the Amati relation?}


\author{M.G. Bernardini}{
  address={Dipartimento di Fisica, Universit\`a di Roma ``La Sapienza'', Roma, I-00185, Italy}
  ,altaddress={ICRANet and ICRA, Piazzale della Repubblica 10, Pescara, I-65122, Italy}
}

\author{C.L. Bianco}{
  address={Dipartimento di Fisica, Universit\`a di Roma ``La Sapienza'', Roma, I-00185, Italy}
  ,altaddress={ICRANet and ICRA, Piazzale della Repubblica 10, Pescara, I-65122, Italy}
}

\author{L. Caito}{
  address={Dipartimento di Fisica, Universit\`a di Roma ``La Sapienza'', Roma, I-00185, Italy}
  ,altaddress={ICRANet and ICRA, Piazzale della Repubblica 10, Pescara, I-65122, Italy}
}

\author{M.G. Dainotti}{
  address={Dipartimento di Fisica, Universit\`a di Roma ``La Sapienza'', Roma, I-00185, Italy}
  ,altaddress={ICRANet and ICRA, Piazzale della Repubblica 10, Pescara, I-65122, Italy}
}

\author{R. Guida}{
  address={Dipartimento di Fisica, Universit\`a di Roma ``La Sapienza'', Roma, I-00185, Italy}
  ,altaddress={ICRANet and ICRA, Piazzale della Repubblica 10, Pescara, I-65122, Italy}
}

\author{R. Ruffini}{
  address={Dipartimento di Fisica, Universit\`a di Roma ``La Sapienza'', Roma, I-00185, Italy}
  ,altaddress={ICRANet and ICRA, Piazzale della Repubblica 10, Pescara, I-65122, Italy}
}

\begin{abstract}
On the basis of the recent understanding of GRB050315 and GRB060218, we return to GRB970228, the first Gamma-Ray Burst (GRB) with detected afterglow. We proposed it as the prototype for a new class of GRBs with ``an occasional softer extended emission lasting tenths of seconds after an initial spikelike emission''. Detailed theoretical computation of the GRB970228 light curves in selected energy bands for the prompt emission are presented and compared with observational \emph{Beppo}SAX data. From our analysis we conclude that GRB970228 and likely the ones of the above mentioned new class of GRBs are ``canonical GRBs'' have only one peculiarity: they exploded in a galactic environment, possibly the halo, with a very low value of CBM density. Here we investigate how GRB970228 unveils another peculiarity of this class of GRBs: they do not fulfill the ``Amati relation''. We provide a theoretical explanation within the fireshell model for the apparent absence of such correlation for the GRBs belonging to this new class.
\end{abstract}

\maketitle

\section{Introduction}\label{intro}

Using the recent understanding of GRB050315 \cite{050315} and GRB060218 \cite{060218}, we point out how GRB970228 can be considered as a prototypical source \cite{970228} useful to explain the properties of a whole class of GRBs characterized by being somehow out from the current GRBs classification \cite{k92,da92} in ``short'' (lasting less than $\sim 2$ s) and ``long'' (lasting more than $\sim 2$ s up to $\sim 1000$ s). 

In the year 2005 there were detected for the first time the X-ray and optical counterparts of two events classified as ``short'': GRB050509B \cite{ge05} by \emph{Swift} and, two months later, GRB050709 \cite{villasenor} by HETE-2. Since then, several ``short'' GRBs' afterglows have been observed, some of them characterized by a soft long bump following the main event, and by a late time evolution similar to the ``long'' GRBs' ones \cite{barthelmy_short}. 

These discoveries motivated Norris \& Bonnell \cite{nb06} to reanalyze the BATSE and HETE II catalogs identifying a new GRB class with ``an occasional softer extended emission lasting tenths of seconds after an initial spikelike emission''. In some cases, ``the strength of the extended emission converts an otherwise short burst into one with a duration that can be tens of seconds, making it appear to be a long burst'' \cite{nb06}, changing the standard GRBs classification in ``short'' and ``long''. Hence, Norris \& Bonnell \cite{nb06} suggested that such a scheme ``is at best misleading''. 

Also GRB060614 \cite{ge06}, indeed, appears to be related to this class of GRBs. In fact, it ``reveals a first short, hard-spectrum episode of emission (lasting $5$ s) followed by an extended and somewhat softer episode (lasting $\sim 100$ s)'': a ``two-component emission structure'' which is ``similar'' to the one analyzed by Norris \& Bonnell \cite{nb06}. GRB060614, due to its ``hybrid'' observational properties, ``opens the door on a new GRB classification scheme that straddles both long and short bursts'' \cite{ge06}.

We analyzed GRB970228 \cite{970228} within our ``canonical GRB'' scenario \cite{rlet1,rlet2,XIIBSGC}, where all GRBs are generated by a ``common engine'' (i.e. the gravitational collapse to a black hole) and we proposed it as the prototype for the new class of GRBs comprising GRB060614 and the GRBs analyzed by Norris \& Bonnell \cite{nb06} since it shares the same morphology and observational properties \cite{970228}. With this result we investigated the $E_{p,i}$--$E_{iso}$ correlation, the so-called ``Amati relation'' \cite{aa02}, for GRB970228 in order to provide a theoretical explanation for the apparent absence of such correlation for the GRBs belonging to this new class.

\section{The ``canonical GRB'' scenario}\label{canonical}

We assume that all GRBs, including the ``short'' ones, originate from the gravitational collapse to a black hole \cite{rlet2,XIIBSGC}. The $e^\pm$ plasma created in the process of the black hole formation expands as a spherically symmetric ``fireshell'' with a constant width on the order of $\sim 10^8$ cm in the laboratory frame, i.e. the frame in which the black hole is at rest. We have only two free parameters characterizing the source, namely the total energy $E_{e^\pm}^{tot}$ of the $e^\pm$ plasma and its baryon loading $B\equiv M_Bc^2/E_{e^\pm}^{tot}$, where $M_B$ is the total baryons' mass \cite{rswx00}. They fully determine the optically thick acceleration phase of the fireshell, which lasts until the transparency condition is reached and the Proper-GRB (P-GRB) is emitted \cite{rlet2}. The afterglow emission then starts due to the collision between the remaining optically thin fireshell and the CircumBurst Medium (CBM) and it clearly depends on the parameters describing the effective CBM distribution: its density $n_{cbm}$ and the ratio ${\cal R}\equiv A_{eff}/A_{vis}$ between the effective emitting area of the fireshell $A_{eff}$ and its total visible area $A_{vis}$ \cite{rlet02,spectr1,fil}. The afterglow luminosity consists of a rising branch, a peak, and a decaying tail \cite{rlet2}.

\begin{figure}
  \includegraphics[height=0.25\textwidth]{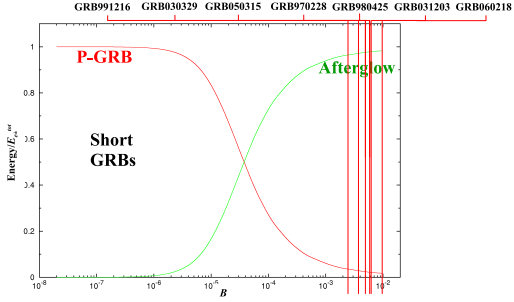}
  \caption{The energy radiated in the P-GRB and in the afterglow, in units of $E_{e^\pm}^{tot}$, are plotted as functions of $B$. Also represented are the values of the $B$ parameter computed for all the GRBs we analyzed. The ``genuine'' short GRBs have a P-GRB predominant over the afterglow: they occur for $B \leq 10^{-5}$ \citep{rlet2}.}
\label{figX}
\end{figure}

Hence, within the fireshell model we define a ``canonical GRB'' light curve with two different components. The first one is the P-GRB and the second one is the afterglow \cite{rlet2,XIIBSGC}. The ratio between the total time-integrated luminosities of the P-GRB and of the afterglow (namely, their total energies) as well as the temporal separation between their peaks are functions of the $B$ parameter \cite{rlet2} and are the crucial quantities for the identification of GRBs' nature. When the P-GRB is the leading contribution to the emission and the afterglow is negligible we have a ``genuine'' short GRB \cite{rlet2,970228}. They correspond to the cases where $B \leq 10^{-5}$ (see Fig.~\ref{figX}). In the limit $B \to 0$ the afterglow vanishes (see Fig.~\ref{figX}). In the opposite limit, for $10^{-4} \leq B \leq 10^{-2}$ the afterglow component is predominant and this is indeed the case of most of the GRBs we have recently examined, including GRB970228 (see Fig.~\ref{figX}). Still, this case presents two distinct possibilities: the afterglow peak luminosity can be either larger or smaller than the P-GRB one. The simultaneous occurrence of an afterglow with total time-integrated luminosity larger than the P-GRB one, but with a smaller peak luminosity, is indeed explainable in terms of a peculiarly small average value of the CBM density and not due to the intrinsic nature of the source. In this sense, GRBs belonging to this class are only ``fake'' short GRBs as in the case of GRB970228 \cite{970228}. We identify the initial spikelike emission with the P-GRB, and the late soft bump with the peak of the afterglow. 

\section{GRB970228 observational properties}\label{observ}

GRB970228 was detected by the Gamma-Ray Burst Monitor (GRBM, $40$--$700$ keV) and Wide Field Cameras (WFC, $2$--$26$ keV) on board \emph{Beppo}SAX on February $28.123620$ UT \cite{fr98}. The burst prompt emission is characterized by an initial $5$ s strong pulse followed, after $30$ s, by a set of three additional pulses of decreasing intensity \cite{fr98}. Eight hours after the initial detection, the NFIs on board \emph{Beppo}SAX were pointed at the burst location for a first target of opportunity observation and a new X-ray source was detected in the GRB error box: this is the first ``afterglow'' ever detected \cite{costa}. A fading optical transient has been identified in a position consistent with the X-ray transient \cite{vp}, coincident with a faint galaxy with redshift $z=0.695$ \cite{bloom01}. Further observations by the Hubble Space Telescope clearly showed that the optical counterpart was located in the outskirts of a late-type galaxy with an irregular morphology \cite{sau97}.

\begin{figure}
  \includegraphics[height=0.35\textheight]{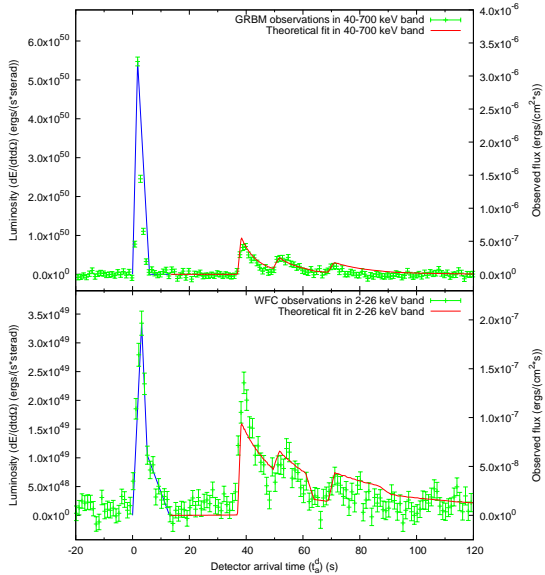}
  \caption{\emph{Beppo}SAX GRBM ($40$--$700$ keV, above) and WFC ($2$--$26$ keV, below) light curves (green points) compared with the theoretical ones (red lines). The onset of the afterglow coincides with the end of the P-GRB (represented qualitatively by the blue lines).}
\label{970228_fit_prompt}
\end{figure}

The \emph{Beppo}SAX observations of GRB970228 prompt emission revealed a discontinuity in the spectral index between the end of the first pulse and the beginning of the three additional ones \cite{costa,fr98,fr00}. The spectrum during the first $3$ s of the second pulse is significantly harder than during the last part of the first pulse \cite{fr98,fr00}, while the spectrum of the last three pulses appear to be consistent with the late X-ray afterglow \cite{fr98,fr00}. This was soon recognized by Frontera et al. \cite{fr98,fr00} as pointing to an emission mechanism producing the X-ray afterglow already taking place after the first pulse.

\section{The analysis of GRB970228 prompt emission}\label{theo}

In Fig.~\ref{970228_fit_prompt} we present the theoretical fit of \emph{Beppo}SAX GRBM ($40$--$700$ keV) and WFC ($2$--$26$ keV) light curves of GRB970228 prompt emission \cite{fr98,970228}. Within our ``canonical GRB'' scenario we identify the first main pulse with the P-GRB and the three additional pulses with the afterglow peak emission, consistently with the above mentioned observations by Costa et al. \cite{costa} and Frontera et al. \cite{fr98}. Such last three pulses have been reproduced assuming three overdense spherical CBM regions \cite{970228} with a very good agreement (see Fig.~\ref{970228_fit_prompt}).

We therefore obtain for the two parameters characterizing the source in our model $E_{e^\pm}^{tot}=1.45\times 10^{54}$ erg and $B = 5.0\times 10^{-3}$ \cite{970228}. This implies an initial $e^\pm$ plasma created between the radii $r_1 = 3.52\times10^7$ cm and $r_2 = 4.87\times10^8$ cm with a total number of $e^{\pm}$ pairs $N_{e^\pm} = 1.6\times 10^{59}$ and an initial temperature $T = 1.7$ MeV \cite{970228}. The theoretically estimated total isotropic energy emitted in the P-GRB is $E_{P-GRB}=1.1\% E_{e^\pm}^{tot}=1.54 \times 10^{52}$ erg \cite{970228}, in excellent agreement with the one observed in the first main pulse ($E_{P-GRB}^{obs} \sim 1.5 \times 10^{52}$ erg in $2-700$ keV energy band, see Fig.~\ref{970228_fit_prompt}), as expected due to their identification \cite{970228}. After the transparency point at $r_0 = 4.37\times 10^{14}$ cm from the progenitor, the initial Lorentz gamma factor of the fireshell is $\gamma_0 = 199$ \cite{970228}. On average, during the afterglow peak emission phase we have for the CBM $\langle {\cal R} \rangle = 1.5\times 10^{-7}$ and $\langle n_{cbm} \rangle = 9.5\times 10^{-4}$ particles/cm$^3$ \cite{970228}. 

We can state that GRB970228 is a ``canonical GRB'' with a large value of the baryon loading. The difference with e.g. GRB050315 \cite{050315} or GRB060218 \cite{060218} is the low average value of the CBM density $\langle n_{cbm} \rangle \sim 10^{-3}$ particles/cm$^3$ which deflates the afterglow peak luminosity. Hence, the predominance of the P-GRB, coincident with the initial spikelike emission, over the afterglow is just apparent: $98.9\%$ of the total time-integrated luminosity is indeed in the afterglow component. Such a low average CBM density is consistent with the occurrence of GRB970228 in the galactic halo of its host galaxy \cite{sau97,vp}, where lower CBM densities have to be expected \cite{panaitescu06}.

\begin{figure}
  \includegraphics[height=0.3\textwidth]{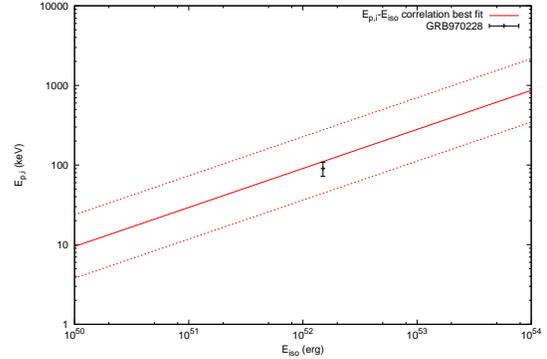}
  \caption{The estimated values for $E_{p,i}$ and $E_{iso}$ obtained by our analysis (black dot) compared with the ``Amati relation'' \cite{aa02}: the solid line is the best fitting power law \cite{amati06} and the dashed lines delimit the region corresponding to a vertical logarithmic deviation of $0.4$ \cite{amati06}. The uncertainty in the theoretical estimated value for $E_{p,i}$ has been assumed conservatively as $20\%$.}
\label{amati}
\end{figure}

\section{GRB970228 and the Amati relation}\label{amati_rel}

We turn now to the ``Amati relation'' \cite{aa02,amati06} between the isotropic equivalent energy emitted in the prompt emission $E_{iso}$ and the peak energy of the corresponding time-integrated spectrum $E_{p,i}$ in the source rest frame. It has been shown by Amati et al. \cite{aa02,amati06} that this correlation holds for almost all the ``long'' GRBs which have a redshift and an $E_{p,i}$ measured, but not for the ones classified as ``short'' \cite{amati06}. If we focus on the ``fake'' short GRBs, namely the GRBs belonging to this new class, at least in one case (GRB050724 \cite{campana_short}) it has been shown that the correlation is recovered if also the extended emission is considered \cite{amatiIK}. 

It clearly follows from our treatment that for the ``canonical GRBs'' with large values of the baryon loading and high $\left\langle n_{cbm}\right\rangle$, which presumably are most of the GRBs for which the correlation holds, the leading contribution to the prompt emission is the afterglow peak emission. The case of the ``fake'' short GRBs is completely different: it is crucial to consider separately the two components since the P-GRB contribution to the prompt emission in this case is significant.

To test this scenario, we evaluated from our fit of GRB970228 $E_{iso}$ and $E_{p,i}$ only for the afterglow peak emission component, i.e. from $t_a^d= 37$ s to $t_a^d= 81.6$ s. We found an isotropic energy emitted in the $2$--$400$ keV energy band $E_{iso}=1.5 \times 10^{52}$ erg, and $E_{p,i}=90.3$ keV. As it is clearly shown in Fig. \ref{amati}, the sole afterglow component of GRB970228 prompt emission is in perfect agreement with the Amati relation. If this behavior is confirmed for other GRBs belonging to this new class, this will enforce our identification of the ``fake'' short GRBs. This result will also provide a theoretical explanation for the the apparent absence of such correlation for the initial spikelike component in the different nature of the P-GRB.

\section{Conclusions}

We conclude that GRB970228 is a ``canonical GRB'' with a large value of the baryon loading \cite{970228}. The difference with e.g. GRB050315 \cite{050315} is the low average value of the CBM density $\langle n_{cbm} \rangle \sim 10^{-3}$ particles/cm$^3$ which deflates the afterglow peak luminosity \cite{970228}. Hence, the predominance of the P-GRB, coincident with the initial spikelike emission, over the afterglow is just apparent. Such a low average CBM density is consistent with the occurrence of GRB970228 in the galactic halo of its host galaxy \cite{sau97,vp}, where lower CBM densities have to be expected \cite{panaitescu06}.

We already proposed GRB970228 as the prototype \cite{970228} for the new class of ``fake'' short GRBs, comprising GRB060614 and the GRBs analyzed by Norris \& Bonnell \cite{nb06}. Most of the sources of this class appear indeed not to be related to the collapse of massive stars, to be in the outskirts of their host galaxies \cite{fox_short} and a consistent fraction of them are in galaxy clusters with CBM densities $\langle n_{cbm} \rangle \sim 10^{-3}$ particles/cm$^3$ \cite{la03,ba07}. This suggests a spiraling out binary nature of their progenitor systems \cite{KMG11} made of neutron stars and/or white dwarfs leading to a black hole formation.

In order to test our interpretation, we verified the applicability of the Amati relation to the sole afterglow component in GRB970228 prompt emission, in analogy with what happens for some of the GRBs belonging to this new class. In fact it has been shown by Amati et al. \cite{amati06,amatiIK} that the ``fake'' short GRBs do not fulfill the $E_{p,i}$--$E_{iso}$ correlation when the sole spiklike emission is considered, while they do if the long soft bump is included. Since the spikelike emission and the soft bump contributions are comparable, it is natural to expect that the soft bump alone will fulfill the correlation as well.

Within our ``canonical GRB'' scenario the sharp distinction between the P-GRB and the afterglow provide a natural explanation for the observational features of the two contributions. We naturally explain the hardness and the absence of spectral lag in the initial spikelike emission with the physics of the P-GRB originating from the gravitational collapse leading to the black hole formation. The hard-to-soft behavior in the afterglow is also naturally explained by the physics of the relativistic fireshell interacting with the CBM, clearly evidenced in GRB031203 \citep{031203} and in GRB050315 \citep{050315}. Therefore, we expect naturally that the $E_{p,i}$--$E_{iso}$ correlation holds only for the afterglow component and not for the P-GRB. Actually we find that the correlation is recovered for the afterglow peak emission of GRB970228.

In the original work by Amati \cite{aa02,amati06} only the prompt emission is considered and not the late afterglow one. In our theoretical approach the afterglow peak emission contributes to the prompt emission and continues up to the latest GRB emission. Hence, the meaningful procedure within our model to recover the Amati relation is to look at a correlation between the total isotropic energy and the peak of the time-integrated spectrum of the whole afterglow. A first attempt to obtain such a correlation has already been performed using GRB050315 as a template, giving very satisfactory results \cite{amati_rob}.

\end{document}